\documentclass[12pt,cite,a4paper]{article}

\usepackage{amsmath}
\usepackage{amssymb}
\begin{document}

\begin{titlepage}
\begin{flushright}
IFUP-TH/2006-41\\
\end{flushright}
~

\vskip .8truecm
\begin{center}
\Large\bf Non invariant zeta-function regularization in
quantum Liouville theory 
\footnote{Work supported in part by M.I.U.R.}
\end{center}

\vskip 1truecm
\begin{center}
{Pietro Menotti} \\
{\small\it Dipartimento di Fisica dell'Universit{\`a}, Pisa 56100,
Italy and}\\
{\small\it INFN, Sezione di Pisa}\\
{\small\it e-mail: menotti@df.unipi.it}\\
\end{center}
\vskip .8truecm

\begin{abstract}

We consider two possible zeta-function regularization schemes of quantum
Liouville theory. One refers to the Laplace-Beltrami operator
covariant under conformal transformations, the other to the naive non
invariant operator. The first produces an
invariant regularization which however does not give rise to a theory
invariant under the full conformal group. The other is equivalent to
the regularization proposed by Zamolodchikov and Zamolodchikov and
gives rise to a theory invariant under the full conformal group.

\end{abstract}
\bigskip

%
%
%
%

\vskip 1truecm

\end{titlepage}

Quantum Liouville theory has been the subject of intense study following
different lines of attack.  While the bootstrap \cite{DO,teschner,ZZsphere,
ZZpseudosphere, FZZ} starts from
the requirement of obtaining a theory invariant under the full
infinite dimensional conformal group, the more conventional field
theory techniques like the hamiltonian and the functional approaches
depend in a critical way on the regularization scheme adopted. In the
hamiltonian treatment \cite{CT} 
for the theory compactified on a circle the
normal ordering regularization gives rise to a theory invariant under
the full infinite dimensional conformal group. 

It came somewhat of a surprise that in the functional approach the
regularization which realizes the full conformal invariance is the non
invariant regularization introduced by Zamolodchikov and Zamolodchikov
(ZZ) \cite{ZZpseudosphere}. In \cite{MV} it was shown that such a
regularization provides the correct quantum dimensions to the vertex
functions on the sphere at least to two loops while in \cite{MT3-4} it
was shown that such a result holds true to all order perturbation
theory on the pseudosphere. Here we 
consider the approach in which the determinant of a
non covariant operator is computed in the framework of the zeta
function regularization and show that this procedure is equivalent to
the non invariant regularization of the Green function at coincident points
proposed by ZZ \cite{ZZpseudosphere}, and extensively used in 
\cite{MV,MT3-4,MT1-2,MT5}. For definiteness we shall refer to the case
of sphere topology.

The complete action is given by
$
S_L[\varphi_B,\chi] = S_{cl}[\varphi_B]+ S_q[\varphi_B,\chi]
$
where  \cite{MV} 
\begin{equation}\label{classicalaction}
S_{cl}[\varphi_B] = \lim_{\stackrel{\varepsilon_n \rightarrow 0}{R
\rightarrow 
\infty}}\frac{1}{b^2}\left[
\frac{1}{8\pi}\int_\Gamma \left(\frac{1}{2}(\partial_a\varphi_B)^2 +8\pi\mu
b^2e^{\varphi_B}\right)d^2 z \right.
\end{equation}
$$
-\sum_{n=1}^{N} \left(\eta_n \,
			\frac{1}{4\pi i}\oint_{\partial\Gamma_{n}}
			\varphi_B \,  
			(\frac{dz}{z-z_n}-\frac{d\bar z}{\bar z-\bar z_n})
+\eta_n^2 \log\varepsilon_n^2\right) \nonumber\\
\left.+\frac{1}{4\pi i}\oint_{\partial\Gamma_{R}} \varphi_B \, \left(
			\frac{dz}{z} - \frac{d\bar{z}}{\bar{z}}
			\right) + \, \log R^2\right] 
$$
and
\begin{eqnarray}\label{quantumaction}
S_q[\varphi_B,\chi] &= & \lim_{\stackrel{\varepsilon_n \rightarrow 0}{R
\rightarrow 
\infty}} \left[\frac{1}{4\pi}\int_\Gamma\left((\partial_a
\chi)^2 + 4\pi \mu e^{\varphi_B}(e^{2b\chi}-1-2b\chi)\right)d^2z \right.
\nonumber \\ 
&+& \left. (2+b^2)\log R^2 +
\frac{1}{4\pi
  i}\oint_{\partial\Gamma_R}\varphi_B\left(\frac{dz}{z}-\frac{d\bar z}{\bar
  z}\right)+ 
\frac{b}{2\pi i}\oint_{\partial\Gamma_R}\chi \left(\frac{dz}{z}-\frac{d\bar
  z}{\bar z}\right)\right].
\end{eqnarray}
In eq.(\ref{quantumaction}) $\Gamma$ is a disk of radius $R$ from 
which disks of radius
$\varepsilon_n$ around the singularities have been removed.

We recall that $S_{cl}$ is $O(1/b^2)$ 
while the first integral appearing in the quantum action
(\ref{quantumaction}) can be expanded as 
\begin{equation}
\frac{1}{4\pi}\int_\Gamma\left((\partial_a
\chi)^2 + 8\pi \mu b^2 e^{\varphi_B}\chi^2 + 8\pi \mu b^2
e^{\varphi_B}(\frac{4b \chi^3}{3!}+\frac{8 b^2\chi^4}{4!}+\dots\right) d^2z . 
\end{equation}
The quantum $n$-point function is given by
\begin{equation}\label{npoint} 
\left< V_{\alpha_1}(z_1)  V_{\alpha_2}(z_2) \dots
V_{\alpha_n}(z_n)\right>= e^{-S_{cl}[\varphi_B]} \int {\cal D}[\chi]~ 
e^{-S_q} 
\end{equation}
where the $\varphi_B$ appearing in the classical and quantum actions 
is the solution of the classical Liouville equation in
presence of $n$ sources
\begin{equation}\label{Liouville}
-\Delta\varphi_B + 8\pi\mu b^2 e^{\varphi_B(z)}
=8\pi\sum_{i=1}^n\eta_i\delta^2(z-z_i) 
\end{equation}
where $\eta_i = b \alpha_i$ and the vertex functions are given by
\begin{equation}
V_\alpha(z)=e^{2\alpha \phi(z)}=e^{\eta
\varphi(z)/b^2};~~~~\varphi=2b\phi=\varphi_B+2b\chi. 
\end{equation}
We recall that the action (\ref{classicalaction}) 
ascribes to the vertex function
$V_\alpha(z)$ the semiclassical dimension
$\Delta_{sc}(\alpha)=\alpha(1/b-\alpha)$ \cite{takh}.

In performing the perturbative expansion in $b$ we have to keep
$\eta_1,\dots \eta_n$ constant \cite{ZZsphere}. 
The one loop contribution to the $n$-point function is given by
\begin{equation}\label{determinant}
K^{-\frac{1}{2}}= \int {\cal D}[\chi]~ 
e^{-\frac{1}{2}\int \chi(z) D \chi(z)d^2z}
\end{equation}
where
\begin{equation}\label{Doperator}
D=-\frac{2}{\pi}\partial_z\partial_{\bar z}+4\mu b^2 e^{\varphi_B}\equiv
-\frac{1}{2\pi}\Delta+m^2e^{\varphi_B}.
\end{equation}
The usual invariant zeta-function technique \cite{zeta} 
for the computation of the functional determinant $K$
consists in writing
\begin{equation}
\int\chi(z)D\chi(z)~ d^2z=
\int\chi(z)\left(-\frac{1}{2\pi}\Delta_{LB}+m^2\right)\chi(z)~ d\rho(z)
\end{equation}
being $d\rho(z) = e^{\varphi_B(z)}d^2z$ the conformal invariant
measure and 
\begin{equation}
\Delta_{LB} =e^{-\varphi_B}\Delta
\end{equation}
the covariant Laplace-Beltrami operator on the background
$\varphi_B(z)$ generated by the $n$ charges. The determinant of the
elliptic operator $-\frac{1}{2\pi}\Delta_{LB}+m^2$ is defined through the
zeta function
\begin{equation}\label{zetasum}
\zeta(s) = \sum_{i=1}^\infty \lambda_i^{-s}
\end{equation}
being $\lambda_i$ the eigenvalues of the operator $H$
\begin{equation}\label{Hdefinition}
H\varphi_i= 
\lambda_i\varphi_i~~~~{\rm where}~~~~ H= -\frac{1}{2\pi}\Delta_{LB}+m^2. 
\end{equation}
For an elliptic operator the sum (\ref{zetasum}) converges for ${\rm Re}~ s$ 
sufficiently
large and positive and the determinant is defined by analytic continuation
as
\begin{equation}
-\log({\rm Det} H) =\zeta'(0).
\end{equation}
Such a value is usually computed by the heat kernel technique
\cite{dewitt,BV}, which
we shall also employ in the following. The great advantage of the
zeta-function regularization is to provide an invariant regularization
scheme as the eigenvalues $\lambda_i$ are invariant under conformal
transformations ($SL(2C)$ for the sphere, $U(1,1)$ for the
pseudosphere).

Associated to the operator $H=-\frac{1}{2\pi}\Delta_{LB}+m^2$ we can
consider the Green
function
\begin{equation}
H G(z,z') = \delta^2(z-z') 
e^{-\varphi_B(z')}\equiv \delta_I(z,z')
\end{equation}
where $\delta_I(z,z')$ is the invariant delta function.

Alternatively one can consider the elliptic operator $D$ defined in
(\ref{Doperator}) and its determinant generated by the zeta function
\begin{equation}
\zeta_D(s) = \sum_{i=1}^\infty \mu_i^{-s}
\end{equation}
being $\mu_i$ the eigenvalues
\begin{equation}
D\psi_i =(-\frac{1}{2\pi}\Delta+m^2 ~e^{\varphi_B(z)})\psi_i(z)=\mu_i\psi_i(z)
\end{equation}
and thus
\begin{equation}\label{Ddeterminant}
-\log({\rm Det}~D)=\zeta'_D(0).
\end{equation}
The Green function for the operator $D$ is defined by
\begin{equation}\label{DGreen}
D g(z,z')=\delta(z-z')
\end{equation}
and multiplying by $e^{-\varphi_B(z)}$ we see that
$G(z,z') = g(z,z')$.
The determinant (\ref{Ddeterminant}), 
being $D$ non covariant is not an invariant
under conformal transformations.

By using the spectral representation we can also write
\begin{equation}
G(z,z') = \sum_i\frac{\varphi_i(z) \varphi_i(z')}{\lambda_i}
\end{equation}
\begin{equation}
g(z,z') = \sum_i\frac{\psi_i(z) \psi_i(z')}{\mu_i}
\end{equation}
where the $\varphi_i(z)$ and $\psi_i(z)$ are normalized by
\begin{equation}
\int |\varphi_i(z)|^2 d\rho(z)=1;~~~~{\rm and}~~~~\int |\psi_i(z)|^2 d^2z=1
\end{equation}
being $d\rho(z)$ the invariant measure $e^{\varphi_B}(z) d^2z$.
We shall exploit in both cases the heat kernel technique. We have
\begin{equation}\label{Hzetafunction}
\zeta(s)= \frac{1}{\Gamma(s)}\int_0^\infty dt ~t^{s-1}{\rm
Tr}(e^{-tH})
\end{equation}
and
\begin{equation}\label{Dzetafunction}
\zeta_D(s)= \frac{1}{\Gamma(s)}\int_0^\infty dt ~t^{s-1}{\rm
Tr}(e^{-tD}). 
\end{equation}
We recall that given the short time expansion
\begin{equation}
{\rm Tr}(e^{-t H })=\frac{c_{-1}}{t}+c_{0}+\dots
\end{equation}
we have $\zeta(0)= c_0$ and
\begin{equation}\label{Hdeterminant}
-\log({\rm Det} H) 
\equiv \zeta'(0) = \gamma_E \zeta(0)+
{\rm Finite}_{\varepsilon\rightarrow 0}\int_\varepsilon^\infty
\frac{dt}{t} {\rm Tr}(e^{-tH})
\end{equation}
where
\begin{equation}
{\rm Finite}_{\varepsilon\rightarrow 0}\int_\varepsilon^\infty
\frac{dt}{t} {\rm Tr}(e^{-tH})= \lim_{\varepsilon\rightarrow 0}
\left[\int_\varepsilon^\infty
\frac{dt}{t} {\rm Tr}(e^{-tH}) -\frac{c_{-1}}{\varepsilon}+
c_0\log\varepsilon\right]
\end{equation}
which gives the value of the determinant of the operator $H$ and 
similarly for $D$.

We examine first (\ref{Dzetafunction}) writing 
\begin{equation}
D=-\frac{1}{2\pi}\Delta + V(z)~~~~{\rm with}~~~~V(z)
=m^2 ~e^{\varphi_B(z)}.
\end{equation}
We have
\begin{equation}
\langle z|e^{\frac{\Delta}{2\pi}t}|z'\rangle =
\frac{1}{2t}e^{-\pi\frac{|z-z'|^2}{2t}} 
\end{equation}
where
\begin{equation}
(e^{\frac{\Delta}{2\pi}t}f)(z)=\int
\langle z |e^{\frac{\Delta}{2\pi}t}|z'\rangle d^2z' f(z') 
\end{equation}
and
\begin{equation}\label{shorttimeexp}
e^{-(-\frac{\Delta}{2\pi}+V)t}=e^{\frac{\Delta}{2\pi}t}-
\int_0^t e^{\frac{\Delta}{2\pi}(t-t')}
~V~e^{\frac{\Delta}{2\pi}t'}dt' +\dots  
\end{equation}
so that
\begin{equation}
{\rm Tr}(e^{-(-\frac{\Delta}{2\pi}+V)t}) =
\frac{1}{2t}\int d^2z -\frac{1}{2t}\int_0^t dt'\int V(z)~ d^2z
+ O(t).
\end{equation}
The coefficient of the first term is the infinite volume term; as such
term 
is proportional to $1/t$ it does not contribute to the finite part. In
addition we see that 
\begin{equation}
c_0=\zeta_D(0) =-\frac{1}{2}\int V(z)~ d^2z. 
\end{equation}
We compute
\begin{equation}\label{DetDHK}
-\log({\rm Det} D) \equiv \zeta_D'(0) = \gamma_E \zeta_D(0)+
{\rm Finite}_{\varepsilon\rightarrow 0}\int_\varepsilon^\infty
\frac{dt}{t} {\rm Tr}(e^{-tD})
\end{equation}
by taking a variation of the background
$\varphi_B$ as due e.g. to a change of the strength of the sources.

We have
\begin{equation}
-\delta \log({\rm Det}~D)=\gamma_E\delta \zeta_D(0) - {\rm
Finite}_{\varepsilon \rightarrow 0} \int_\varepsilon^\infty dt~ {\rm
Tr}(\delta Ve^{-tD})=
$$
$$
=\gamma_E\delta \zeta_D(0) - {\rm
Finite}_{t \rightarrow 0} {\rm
Tr}(\delta VD^{-1}e^{-tD})=
$$
$$
=\gamma_E\delta \zeta_D(0) - {\rm
Finite}_{t\rightarrow 0} \int d^2z \,\delta V(z)\,g(z,z')d^2z'
\langle z'|e^{-t D}|z \rangle
\end{equation}
being $g(z,z')$ the Green function defined in (\ref{DGreen}). The
finite part appearing in the above equation  can be computed by
exploiting again the short time expansion (\ref{shorttimeexp}). 

We have
\begin{equation}\label{OVeq}
-\int d^2z ~\delta V(z)~g(z,z')~d^2z'
~\langle z'|e^{-t D}|z \rangle=
-\int d^2z ~\delta V(z)~g(z,z')~d^2z'~
\frac{1}{2t}e^{-\pi\frac {|z-z'|^2}{2t}}+ O(V).
\end{equation}
We shall show in the following that the $O(V)$ term does
not contribute to the finite 
part as it goes to zero for $t\rightarrow 0$. 
With regard to the first term it equals
\begin{equation}
-\int d^2z ~\delta V(z)\left(-\frac{1}{2}\log|z-z'|^2+ g_F(z)+ o(|z-z'|
\right)d^2z'~
\frac{1}{2t}e^{-\pi\frac {|z-z'|^2}{2t}}
\end{equation}
which for $t\rightarrow 0$ goes over to
\begin{equation}\label{separationgF}
-\int \delta V(z) ~g_F(z)~ d^2z+\int d^2z ~\delta V(z)~\frac{1}{2}\log|z-z'|^2
~d^2z'~\frac{1}{2t}e^{-\pi\frac {|z-z'|^2}{2t}}
\end{equation}
and we have written
\begin{equation}
g(z,z')=-\frac{1}{2}\log|z-z'|^2+g_F(z) +o(|z-z'|).
\end{equation}
Thus we have to compute the finite part of the second term in 
eq.(\ref{separationgF}).
This is easily done with the result
\begin{equation}
{\rm Finite}_{t\rightarrow
0}\frac{1}{2}(\log t-\log\frac{\pi}{2}-\gamma_E)\int
d^2z~\delta V(z)
=-\frac{1}{2}(\log\frac{\pi}{2}+\gamma_E)\int
d^2z~\delta V(z). 
\end{equation}
We must now examine in eq.(\ref{OVeq}) the $O(V)$ term 
\begin{equation}\label{singintegral}
I_V= \int d^2z~ \delta V(z) g(z,z') d^2z'\int_0^t
\frac{1}{2(t-t')}e^{-\frac{\pi|z'-z^{''}|^2}{2(t-t')}}V(z^{''})
~d^2z^{''}\frac{1}{2t'}e^{-\frac{\pi|z^{''}-z|^2}{2t'}}dt'.
\end{equation}
We have that $\delta V(z)$ is integrable and also 
\begin{equation}
\int d^2z~ |\delta V(z) \log|z-z'|^2| < \infty.
\end{equation}
At this stage were it $V(z)$ bounded i.e. $|V(z)|<V_M$ it would follow 
immediately that 
\begin{equation}\label{firstbound}
|I_V| < t~V_M ~\left( g_M + c +\frac{1}{2}\log(\frac{\pi}{2t})\right)
~\int \delta
V(z) d^2 z
\end{equation}
being 
\begin{equation}
g_M =\sup_{|z-z'|<1}|g(z,z')+\log|z-z'|~| +\sup_{|z-z'|>1}|g(z,z')|
\end{equation}
and $c$ a numerical constant. The r.h.s. of eq.(\ref{firstbound})
vanishes for $t\rightarrow 0$. 
In the case at hand $V(z)=m^2e^{\varphi_B}$ with
$\varphi_B$ satisfying eq.(\ref{Liouville}) is not
bounded but has locally integrable singularities. Such a situation can
be dealt with as follows. Isolate the singularities with disks of given
radius around them. For the contribution outside these disks the above
reasoning applies. For the disk contributions we can use the following
bound. Let the singularity of $e^{\varphi_B(z)}$ be located in zero and
given by ${\rm const}~ |z|^{-2\gamma}$ where due to the local integrability
of the area $\gamma<1$. 
$$
\int_0^t dt'
\int\frac{1}{2(t-t')}e^{-\pi\frac{|z'-z^{''}|^2}{2(t-t')}}|z^{''}|^{-2\gamma} 
d^2 z^{''} 
\frac{1}{2t'}e^{-\pi\frac{|z^{''}-z|^2}{2t'}} =
$$
$$
=\frac{t^{\gamma-1}}{4}\int_0^t\frac{dt'}{t'(t-t')^\gamma} 
\int e^{-\pi\frac{|z'-z|^2}{2t}} e^{-\frac{u^2}{2}}
\left|u +2(\frac{z'}{t-t'}+\frac{z}{t'})\sqrt{\frac{t'(t-t')}{t}} 
\right|^{-2\gamma}
d^2 u <
$$
$$
=\frac{t^{\gamma-1}}{4}\int_0^t\frac{dt'}{t'(t-t')^\gamma} 
\int e^{-\pi\frac{|z'-z|^2}{2t}} e^{-\frac{u^2}{2}}
\left|u \right|^{-2\gamma}
d^2 u <
$$
\begin{equation}
<2^{\gamma-2}\pi^{\gamma+1/2}\frac{\Gamma^2(1-\gamma)}{\Gamma(3/2-\gamma)}
t^{-\gamma}e^{-\pi\frac{|z'-z|^2}{2t}}.
\end{equation}
As $\gamma<1$ repeating the argument which leads to
eq.(\ref{firstbound}) 
we have that the integral (\ref{singintegral}) goes to zero for
$t\rightarrow 0$.

Putting all contributions together we find 
\begin{equation}\label{noninvresult}
-\delta\zeta'_D(0)= \delta \log({\rm Det}D) = 
\int d^2z ~\delta V(z)  \left(g_F(z)
+\frac{1}{2}\log\frac{\pi}{2}+\gamma_E\right).
\end{equation}
Thus we have for the variation of the determinant the same result
which is obtained formally from
\begin{equation}
\delta \log({\rm Det}D)^{-\frac{1}{2}} = -\frac{1}{2}
\int d^2z ~\delta V(z) g(z,z)
\end{equation}
where one replaces the divergent quantity $g(z,z)$ by the regularized
value \cite{ZZpseudosphere}
\begin{equation}
g(z,z) = \lim_{z'\rightarrow z}(g(z,z')+\frac{1}{2}\log|z-z'|^2)+C=
g_F(z)+C
\end{equation}
with $C=\frac{1}{2}\log(\pi/2)+\gamma_E$. The contribution
(\ref{noninvresult}) changes the dimensions of the vertex field
$V_\alpha(z)$ from the semiclassical value $\Delta_{sc}(\alpha) =
\alpha(1/b-\alpha)$ to the quantum dimension
$\Delta(\alpha)=\alpha(1/b+b-\alpha)$~
\cite{ZZpseudosphere,MV,MT3-4,MT1-2,MT5}. 
At the perturbative level on the
pseudosphere ZZ \cite{ZZpseudosphere} chose $C=0$. 
Such a constant can be absorbed in a constant shift in the Liouville field.
This can be shown for the contribution of the quantum determinant to
all order in $\eta_j = b\alpha_j$ for the $n$-point function of 
eq.(\ref{npoint}). In fact the
background field solves eq.(\ref{Liouville}); taking the
derivative w.r.t. $\eta_j$ and integrating over the whole plane we
have
\begin{equation}\label{derintegral}
\int d^2z\frac{\partial (\mu b^2 e^{\varphi_B})}
{\partial \eta_j}d^2 z = 1
\end{equation}
insofar due to the behavior of $\partial (\mu b^2
e^{\varphi_B})/\partial \eta_j$ at infinity only the gradient of the
field $\varphi_B$ around the charges contributes to the boundary
integral and the
integral (\ref{derintegral}) converges. As a result 
integrating back in $\eta_j$,  $({\rm Det}D)^{-1/2}$ gets multiplied by
$e^{-2C\eta_j}=e^{-2Cb\alpha_j}$ corresponding to the shift in the field 
$\phi\rightarrow \phi-bC$. The fact that a change of the regularized
value of the Green function at coincident points is equivalent to the
above mentioned shift in the field $\phi$ can be  shown on the
pseudosphere to all order perturbation theory by exploiting the
identity  which relates the tadpole graph and the
simple loop \cite{ZZpseudosphere}
\begin{equation}\label{tadpolecanc}
-4b^2\mu \int \hat g(z,z')d\rho(z') \hat g(z',z) + \hat g(z,z)=\frac{1}{2}
\end{equation}
which holds for the Green function on the background of the
pseudosphere \cite{ZZpseudosphere,DFJ}
\begin{equation}
\hat g(z,z')= -\frac{1}{2}\left(\frac{1+\omega}{1-\omega}\log \omega +2\right)
\end{equation}
with $\omega = |(z-z')/(1-z'\bar z)|^2$
and extending a combinatorial argument developed in \cite{MT3-4}
in connection with the quantum dimensions of the vertex function on the
pseudosphere to all order perturbation theory.

\bigskip

We want now to compare the result (\ref{noninvresult}) 
with the computation of the
determinant of the covariant operator $H$ eq.(\ref{Hdefinition}). 
We recall that in the
eigenvalue equation (\ref{Hdefinition}) in presence of conical 
singularities it may occur that both behaviors of the solution $\varphi_n$
at the singularity 
are square integrable in the invariant metric $d\rho(z)$ \cite{MP} and
this fact gives rise to the problem of the self-adjoint extension of
the operator (\ref{Hdefinition}). 
A standard way for doing that is to regularize
the singularities and then take the limit when the regularization is
removed. The main point however is that the so defined operator has an
invariant spectrum and as such through the zeta- function procedure
gives rise to an invariant value for the functional determinant. 
The derivative w.r.t. $\eta_j$ of the change of 
$\log({\rm Det}H)^{-\frac{1}{2}}$ under dilatations vanishes and the
same happens for the $O(b^0)$ boundary terms of eq.(\ref{quantumaction}).
Thus
at one loop the correction to the semiclassical dimensions $\Delta_{sc}
(\alpha) =
\alpha(1/b-\alpha)$ vanishes. In particular at one loop the
cosmological term $e^{2b\phi}$ maintains the weights
$(\Delta(b),\Delta(b))=(1-b^2,1-b^2)$. Higher order corrections cannot 
bring such weights to
the value $(1,1)$, because the two loop correction is of the form
$f(\eta)~b^2$ \cite{MV} and as we must have for the dimension,
$\Delta(0)=0$ we have $f(0)=0$. But then at two loop we have
$\Delta(b)= 1-b^2+ O(b^4)$ which cannot be identically $1$ in $b$.

For the sake of comparison we shall compute the determinant
of the covariant operator $H$ given in eq.(\ref{Hdefinition}) 
in the regularized case. The related 
zeta-function is given by eq.({\ref{Hzetafunction})
where the trace is
\begin{equation}
{\rm Tr}(e^{-tH})=\sum_n\int d\rho(z) \varphi_n(z) (e^{-tH}
\varphi_n)(z) = 
\end{equation}
$$
\int d\rho(z) \varphi_n(z)
\langle z|e^{-tH}|z'\rangle d^2z' \varphi_n(z')=
\sum_n e^{-t\lambda_n}
$$
and we used the following definition for $\langle
z|e^{-tH}|z'\rangle$
\begin{equation}
(e^{-tH} f)(z) \equiv \int \langle
z|e^{-tH}|z'\rangle d^2z' f(z')
\end{equation}
and  thus it follows
\begin{equation}
{\rm Tr}(e^{-tH})=\int d^2z\langle z|e^{-tH}|z \rangle.
\end{equation}
The determinant of $H$ is given by eq.(\ref{Hdeterminant}).
The value of $\zeta(0)$ in eq.(\ref{Hdeterminant}) is found by standard 
techniques \cite{BV} to be
\begin{equation}
\zeta(0) = c_0 = -\frac{m^2}{2} \int e^{\varphi_B} d^2z +
\frac{1}{24\pi}\int R(z)~ e^{\varphi_B} d^2z
\end{equation}
where $R(z)$ is the curvature related to the metric $e^{\varphi_B} ~i
dz\wedge d\bar z/2$. The first integral apart the multiplicative
constant is the
volume while the second is simply the Gauss-Bonnet term and as such a
topological invariant.

The change of $\zeta'(0)$ under a small change of $\varphi_B$ is given by 
\begin{equation}
\delta \zeta'(0)=\gamma_E \delta \zeta(0) - {\rm
Finite}_{\varepsilon\rightarrow 0}\int_\varepsilon^\infty dt
~d^2 z~\delta\varphi_B(z) e^{-\varphi_B(z)}\langle z|\frac{1}{2\pi}
\Delta e^{-tH}|z\rangle=
\end{equation}
$$
\gamma_E \delta \zeta(0) - \frac{1}{2\pi}{\rm
Finite}_{t \rightarrow 0}\int
d^2 z~\delta\varphi_B(z) e^{-\varphi_B(z)}\Delta G(z,z') d\rho(z')
\langle z'|e^{-t(-e^{-\varphi_B}\frac{\Delta}{2\pi}+m^2)}|z\rangle.
$$
Using
\begin{equation}
-\frac{e^{-\varphi_B(z)}}{2\pi}\Delta G(z,z') = \delta_I(z,z')- m^2~
G(z,z')
\end{equation}
we obtain
\begin{equation}
\delta \zeta'(0) =
\gamma_E\delta \zeta(0)+ 
{\rm Finite}_{t\rightarrow 0}\int d^2z\, \delta\varphi_B(z)
\langle z|e^{-tH}|z\rangle-
\end{equation}
$$
m^2~{\rm Finite}_{t\rightarrow 0}\int d^2z\, \delta\varphi_B(z)\,
G(z,z')\,d\rho(z')\langle z'|e^{-tH}|z\rangle.
$$
Using the same technique as in deriving eq.(\ref{noninvresult}) 
we find
\begin{equation}
\delta \zeta'(0)=
$$
$$
\gamma_E\left(-\frac{m^2}{2}\delta\int e^{\varphi_B(z)}d^2z +
\frac{1}{24\pi}\delta\int R(z)~e^{\varphi_B} d^2z\right)
-\frac{m^2}{2}\int d\rho(z)
\delta\varphi_B(z)+
\end{equation}
$$
\frac{1}{24\pi}\int
\delta\varphi_B(z) R(z)~d\rho(z)
- m^2\int e^{\varphi_B(z)}~d^2z~\delta\varphi_B(z)
(g_F(z)+\frac{1}{2}\varphi_B(z)+\frac{1}{2}
(\log\frac{\pi}{2}+\gamma_E)=
$$
$$
=-\int d\rho(z)\,
\delta\varphi_B(z)(\frac{m^2}{2}-\frac{R(z)}{24\pi})
$$
$$- m^2\int d\rho(z)~\delta\varphi_B(z)
(g_F(z)+\frac{1}{2}\varphi_B(z)+\frac{1}{2}
\log(\frac{\pi}{2})+\gamma_E).
$$
In the first term one recognizes the variation of the conformal
anomaly in presence of the ``mass'' $m$ while the second term is the
same as the result (\ref{noninvresult}) 
except for a different regularization of
the Green function. The additional contribution 
$\varphi_B(z)/2$ can be understood as an invariant
regularization of the Green function at coincident points obtained by
subtracting the divergence $-\frac{1}{2}(\log 2 \sigma(z,z'))$ being 
$2\sigma(z,z')$ the square of the invariant distance which for small
$z-z'$ reduces to $e^{\varphi_B(z)}|z-z'|^2$. Invariant
regularization of this kind for the Green function 
has been considered in the context of
Liouville theory e.g. in \cite{DFJ}; the resulting theory is invariant
only under the group $SO(2,1)$.

In conclusion, the zeta-function regularization \cite{zeta} 
and the related heat
kernel technique \cite{dewitt}  has been introduced in
the literature with the aim of providing an
invariant regulator. Here it has been shown that in quantum
Liouville theory, due to the non invariance of the total action 
$S_L[\varphi_B,\chi] = S_{cl}[\varphi_B]+ S_q[\varphi_B,\chi]$ a
non invariant regularization is necessary and that this is provided by
the zeta-function regularization of a non covariant operator. Such a
situation may well occur in other quantum field theory models.

\section*{Acknowledgments}

I am grateful to Damiano Anselmi for useful discussions.


\begin{thebibliography}{99}


\bibitem{DO} M. Goulian, Miao Li, Phys. Rev. Lett. 66 (1991) 2051;
H. Dorn and H.J. Otto, Nucl. Phys. B429 (1994) 375.

\bibitem{teschner}
J. Teschner, Phys. Lett. B363 (1995) 65; Class.
  Quant. Grav. 18 (2001) R153; Int. J. Mod. Phys.
  A19 S2 (2004) 436; Nucl.Phys.B622 (2002) 309;
  {\it From Liouville Theory to the 
  Quantum Geometry of the  Riemann Surfaces}, hep-th/0308031;
  
\bibitem{ZZsphere}
A.B. Zamolodchikov and Al.B. Zamolodchikov, Nucl. Phys. B477
(1996) 577.

\bibitem{ZZpseudosphere}
A.B. Zamolodchikov and Al.B. Zamolodchikov,
  {\it Liouville Field Theory on a Pseudosphere},
  hep-th/0101152.

\bibitem{FZZ}
V. Fateev, A.B. Zamolodchikov and Al.B. Zamolodchikov,
  {\it Boundary Liouville Field Theory
       I. Boundary State and Boundary Two-point Function},
   hep-th/0001012;\\
J. Teschner,
  {\it Remarks on Liouville theory with boundary}, hep-th/0009138.

\bibitem{CT}
  T.L. Curtright and C.B. Thorn, Phys. Rev. Lett.
  48 (1982) 1309;
  E. Braaten, T.L. Curtright and C.B. Thorn, Phys.
  Rev. Lett. 51 (1983) 19; Ann. Phys. 147 (1983)
  365; G. Jorjadze and G. Weigt, Phys. Lett. B581 (2004) 133.


\bibitem{MV}
P. Menotti and G. Vajente, Nucl. Phys. B709 (2005) 465.
 
\bibitem{MT3-4}
P. Menotti and E. Tonni, Phys. Lett. B633 (2006) 404; JHEP
0606:020,2006

\bibitem{MT1-2}
P. Menotti and E. Tonni, Phys. Lett. B586 (2004) 425;
Nucl. Phys. B707 (2005) 321.
 
\bibitem{MT5}
P. Menotti and E. Tonni, JHEP 0606:022, 2006

\bibitem{takh}
L.A. Takhtajan, Mod. Phys. Lett. A11 (1996) 93.

\bibitem{zeta}
J.S. Dowker and  R. Critchley, Phys. Rev. D 13 (1976) 3224; S. Hawking,
Comm. Math. Phys. 55 (1977) 133.

\bibitem{dewitt} B.S. DeWitt, Dynamical Theory of Groups and Fields,
Gordon and Breach, New York, 1965.


\bibitem{BV}
R. Balian and C. Bloch, Ann. Phys.64 (1971) 271; O. Alvarez,
Nucl. Phys. B216 (1983) 125;
A.O. Barvinski and G.A. Vilkoviski, Phys. Rep. 119 (1985) 1.

\bibitem{MP}
P. Menotti and P.P. Peirano, Phys. Lett. B353 (1995) 444; 
Nucl. Phys. B473 (1996) 426; Nucl.Phys.Proc.Suppl. 57 (1997) 82.

\bibitem{DFJ}
E. D'Hoker, D.Z. Freedman and R. Jackiw, Phys. Rev. D28
(1983) 2583.

\end{thebibliography}
\end{document}